\documentclass{article}

\usepackage{microtype}
\usepackage{graphicx}
\usepackage{subcaption}
\usepackage{booktabs} 

\usepackage{hyperref}

\usepackage[preprint]{icml2026}

\usepackage{amsmath}
\usepackage{amssymb}
\usepackage{mathtools}
\usepackage{amsthm}

\usepackage[capitalize,noabbrev]{cleveref}

\theoremstyle{plain}

\theoremstyle{definition}

\theoremstyle{remark}

\usepackage[textsize=tiny]{todonotes}

\icmltitlerunning{Dynamic Sparse Attention: Access Patterns and Architecture}

\begin{document}

\twocolumn[
  \icmltitle{Dynamic Sparse Attention: Access Patterns and Architecture}




  \begin{icmlauthorlist}
    \icmlauthor{Noam Levy}{yyy}    
  \end{icmlauthorlist}

  \icmlaffiliation{yyy}{Intel Corporation}

  \icmlkeywords{Machine Learning, ICML}

  \vskip 0.3in
]



\printAffiliationsAndNotice{}  

\begin{abstract}
Dynamic sparse attention (DSA) reduces the per-token attention bandwidth by restricting computation to a top-$k$ subset of cached key--value (KV) entries, but its \emph{token-dependent} selection pattern introduces a system-level challenge: the KV working set is fragmented, volatile, and difficult to prefetch, which can translate into poor cache locality and stalled decode throughput. We study these effects by implementing a lightweight indexer for DSA-style selection on multiple open-source backbones and logging per-layer KV indices during autoregressive decoding. Our analysis shows a gap in serving DSA backbones - a potential for a high volume of blocking LL (last level) cache miss events, causing inefficiency; we propose a novel LL cache reservation system to save KV tokens in the LL cache between decode steps, combined with a token-granularity LRU eviction policy, and show on the data we collected how this architecture can benefit serving with DSA implemented on different backbones. Finally, we propose directions for future architectural and algorithmic exploration to improve serving of DSA on modern inference platforms.
\end{abstract}

\section{Introduction}
\label{sec:introduction}

Autoregressive large language models (LLMs) use a variety of techniques to store and attend to context. A significant part of existing work use some variation of self-attention \cite{vaswani2017attention}, which allows a model to attend to any part of its past context.

As we focus specifically on autoregressive inference (the decode stage), attention can be described by:

\begin{equation}
  a = \textnormal{softmax}\big(\frac{q K^T}{\sqrt{d_k}} \big) V
\end{equation}

Where $q$ is the new query vector; the $K$ and $V$ are the collection of all previous key and value projections, and can therefore be cached in memory for efficient lookup. This mechanism, aptly named the \textit{KV cache}, is widely deployed in LLM inference systems \cite{kwon2023pagedattention}. It is easy to see, however, that even with KV caching, the complexity of this single operation scales linearly with the sequence length $T$, and is therefore quadratic in total complexity over the entire sequence.

Even with this quadratic complexity, modern GPUs can often manage the compute load of decoding with sufficient throughput, especially in the decode part and the long context regime, and the bottleneck materializes in the \textbf{memory system}---specifically bandwidth throughput, or goodput (accounting for lost cycles) \cite{dao2023flashattention2}. Table~\ref{tab:h100_util} shows, via a roofline model, that modern LLM backbones in the decode stage of inference are typically bandwidth-limited, while compute remains underutilized. The notable exception of DeepSeek 3.2 is discussed below.

\begin{figure}[t]
  \centering
  \includegraphics[width=0.5\textwidth]{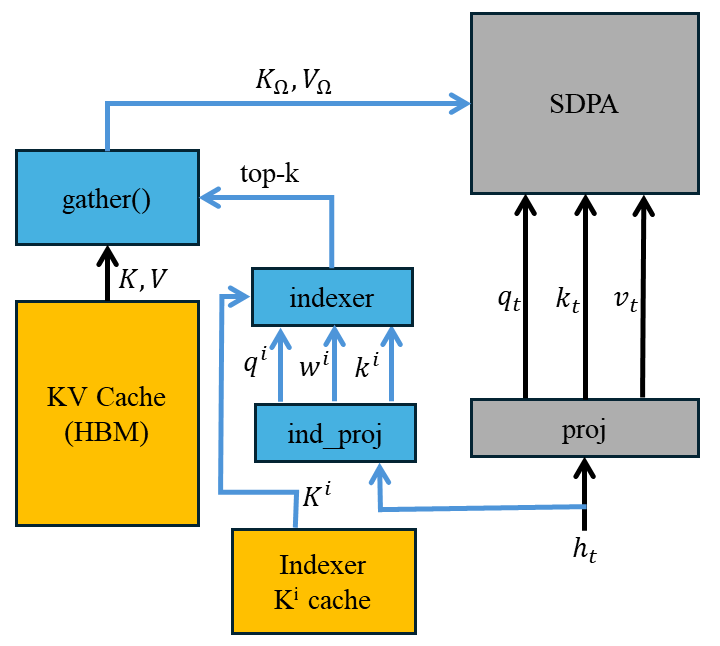}
  \caption{Dynamic Sparse Attention overview; an indexer, using a small number of heads and small projection dimensionality, scores each context token; the top-k score are selected, and a the KV values corresponding are gathered from the KV cache, which is typically stored in HBM memory; the scaled dot product attention is applied only to this subset}
  \label{fig:overview}
\end{figure}

Prior work specifically addresses the potential sparsity of the attention operation: \cite{liu2022dynamic} train a proxy for attention that explicitly utilizes sparsity and allow sparse matrix operators on GPUs to properly exploit it; \cite{dai2023efficient} show an approach that uses statically structured sparsity, potentially benefiting from spatial locality of non-sparse values. These works motivate the discussion, but they do not address the need to bring the selected KV values from memory at decode.

\begin{figure}[t]
  \centering
  \includegraphics[width=0.5\textwidth]{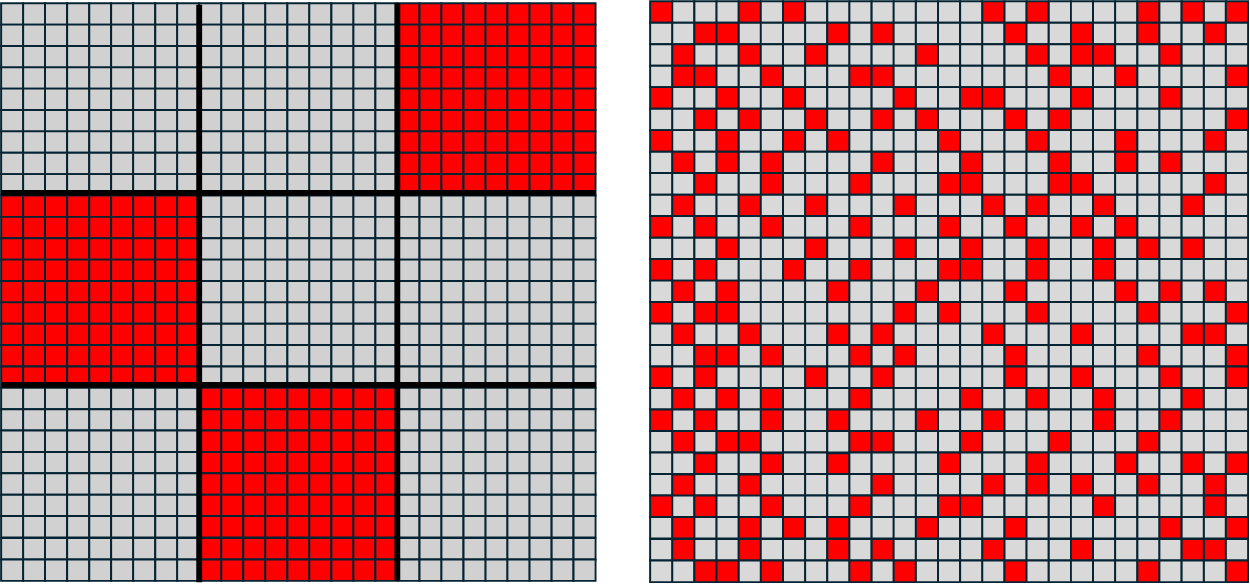}
  \caption{Blockwise sparse attention vs.\ DSA sparse attention; each query may select a completely different top-$k$ KV subset to attend to, eliminating the reuse potential of previous approaches and introducing an implementation challenge.}
  \label{fig:Xattention_vs_DSA}
\end{figure}

One of the promising directions proposed is \textit{dynamic sparse attention}, with a notable implementation in the frontier DeepSeek 3.2 model \cite{liu2025deepseek} (DSA). In this method, illustrated in Figure~\ref{fig:overview}, attention explicitly attends to only a subset of past KV entries, with a fixed budget of $k$ entries. Unlike block-sparse attention \cite{xu2025xattention}, DSA selects a unique subset of KV for each token, making the access pattern much more challenging, as illustrated in Figure~\ref{fig:Xattention_vs_DSA}. 

From a system perspective, the KV cache is also a dominant resource-management challenge for serving: it grows linearly with the number of generated tokens and must be managed dynamically across many concurrent requests. To reduce fragmentation and improve batching, \cite{kwon2023pagedattention} propose \textit{PagedAttention}, a paging-inspired KV cache layout and allocator implemented in the \textsc{vLLM} serving system; the resulting non-contiguous physical layout improves memory utilization but can complicate kernel implementations and impact locality. Complementary work proposes retaining a contiguous virtual KV layout while relying on CUDA virtual memory management for dynamic physical allocation \cite{prabhu2024vattention}, avoiding specialized attention kernels while still mitigating fragmentation. Finally, even when compute-efficient attention kernels reduce data movement within the GPU hierarchy, attention remains fundamentally IO-bound for long contexts; IO-aware kernels such as FlashAttention-2 \cite{dao2023flashattention2} and FlashAttention-3 \cite{shah2024flashattention3} emphasize tiling and overlap of computation with data movement to raise effective bandwidth utilization. These works motivate out approach - as we shift the paradigm of KV memory access with DSA, we must accordingly shift our focus on optimizing the system for it.

With DSA, attention becomes a fixed-complexity operation per token in terms of memory bandwidth because only a fixed amount of data is fetched from the KV cache for each autoregressive token generation step (in the artithmetic of scale, the indexer can rightly be ignored, although it does not follow the fixed-complexity assumption); however, we will show that this subset of past KV pages can be disparate and highly volatile, potentially generating a large volume of L2 cache misses in the decode phase and severely limiting performance due to memory system latencies. Prior work has already shown that a significant miss rate in the GPU L2 cache can introduce an equally significant decode slowdown, even when in all other aspects the system can serve the model efficiently \cite{dong2025accelerating}.

We will show by implementing DSA on several different backbones that its implementation introduces a set of specific and predictable system constraints in inference; we benchmark, profile and model these system constraints and and we propose an architectural modifications to allow efficient implementation - by reserving a portion of the L2 cache to hold DSA KV tokens between decode iterations.

We present an architectural model of this proposed improvement and show that when implemented, it can significantly boost DSA performance by implementing a specialized cache reservation and eviction policy, but without increasing the volume (and therefore die area) of the L2 cache.

\begin{table}[t]
  \centering
  
  \caption{H100 Rack utilization (per GPU) to decode at 100 tokens per second per user with a 64k context and batch size of 8}
  \label{tab:h100_util}
  \begin{tabular}{lccc}
    \toprule
    \textbf{Backbone} & \textbf{N GPUs} & \textbf{HBM BW} & \textbf{Compute} \\
    \midrule
    LLama 3.1 70B & 7 & 95\% & 1.3\% \\
    Llama 4 Maverick & 26 & 99\% & 0.1\%  \\
    GPT-OSS 120B & 8 & 97\% & 0.2\% \\
    QWEN3-480B & 32 & 99\% & 0.2\% \\
    DeepSeek 3.2 & 44 & 93\% & 67\% \\
    \bottomrule
  \end{tabular}
\vspace{-0.75\baselineskip}
\end{table}

\section{Methodology}
\label{sec:methodology}
\subsection{The Lightning Indexer}
\label{sec:lightning-indexer}
In order to evaluate the potential efficacy and pitfalls of the DSA approach, we implemented the \textit{lightning indexer} and trained it into a frozen backbone in 4 open source models from the LlaMa 3 herd \cite{grattafiori2024llama}: LLaMa3.1-70B Instruct, LLaMa 3.1-8B Instruct, LLaMa3.2-3B Instruct, and LLaMa3.2-1B Instruct; we then monkey-patched the attention with a sparse attention supervised by this indexer, following the DeepSeek methodology \cite{liu2025deepseek}.

The indexer is a quick attention-approximation operator to rank the relevance of potential KV entries before they are fetched from memory. Specifically, a $H_i$-headed indexer is implemented as a score $S_{t,s}$ of how the $s$-th KV token is expected to attend the $t$-th output token:

\begin{equation}
  S_{t,s} = \sum_{j=1}^{H_i}w^i_{t,j}\cdot\text{ReLU}\big(q^i_{t,j} \cdot k^i_s\big) 
\end{equation}

Where $k^i$, $q^i$, $w^i$ are projected from the hidden states. The set of $\Omega_t = \textnormal{Top-k}_s \big(S_{t,s}\big)$ is then used for selecting KV values $K_{\Omega_t}$, $V_{\Omega_t}$, and only those are used the attention calculation. The DSA as described in \cite{liu2025deepseek} is used in an MQA setting; since our chosen models use GQA, and to keep the indexer small and efficient, we broadcast $\Omega$ to all KV heads and reuse it.

In our implementation (after a few ablation studies), we selected $D_{\text{indexer}}=64$ (the dimension of each $q^i$ and $k^i$ projection) and $H_i=4$; this introduces $516 \times d_{\text{model}} \times N_{\text{layers}}$ trainable parameters. Here we slightly diverge from the DeepSeek implementation, as the indexer there is combined with the latent-space representation of the attention projections (known as MLA); but to simplify the analysis, and to avoid pretraining the entire model, we did not implement MLA.

In order to train the DSA indexer with otherwise frozen weights, we applied a sigmoid $I=\sigma(S)$ and attempted to directly sparisify $I$ by applying a loss while performing knowledge-distilation from the dense model:

\begin{equation}
  \mathcal{L} = \mathcal{L}_{\text{logits}} + \mathcal{L}_{\text{attn}} +  \mathcal{L}_{\text{sparse}} + \mathcal{L}_{\text{entropy}}
\end{equation}

The main data-term loss $\mathcal{L}_{\text{logits}}$ was $D_{\mathrm{KL}}\!\left(L_{\text{sparse}} \,\|\, L_{\text{dense}}\right)$, aiming to align the final model logits distribution, and a similar $\mathcal{L}_{\text{attn}}$ applied on the attention scores of each model layer. The main sparseness terms were applied directly to $I$, to encourage sparsity and binarization

\begin{align}
  \mathcal{L}_{\text{sparse}} &= \lambda_s \lVert I\rVert_1 \\
  \mathcal{L}_{\text{entropy}}  &= \lambda_e H(I)
\end{align}

For training we used the SlimPajama \cite{cerebras2023slimpajama} dataset, a high quality, curated and de-duped dataset from multiple sources, which proved diverse and clean enough for the relatively easy task of pre-training the indexer. We used sequences from 100 up to 2048 tokens long, with a continuation of 64 tokens, randomly selected from the dataset. We selected a batch size of 4.

Pretraining consisted of 100k training steps using the AdamW optimizer with cosine annealing and an initial learning rate of $3\cdot10^{-4}$; training was performed on an H200 GPU. Weights were frozen from the HuggingFace official release of these models as of Feb 10, 2026. For training only, we cast the internal datatypes to FP32 for numerical stability, but for inference we kept the native BF16.

Since pretraining provided a sufficiently coherent result for the models we trained, we completed the experiments without applying a fine tuning approach to our indexer.

\subsection{Architectural Analysis}
\label{sec:architectural-analysis}
After the model is trained and validated, we build a framework to evaluate the access patterns. The framework follows the methodology of performing a dense prefill stage, and then decoding sequences and collecting statistical data on the KV access patterns.

Each token generated also generated a list of top-k KV references selected to participate in the DSA calculation. Each layer had a different selction of top-k selections and these were logged.

An aggregation script was designed to analyze:
\begin{enumerate}
  \item Size of the \textbf{working set}: how many KV values in total are needed to generate an $N$-token chunk (in our experiments, $N=50$); we report values as a fraction of top-$k$. This indicates the \textit{physical} size of the KV chunk in memory required for "hot" access (assuming efficient paging is applied)   
  \item \textbf{Lookback steps}: how far, on average, the selection reaches into the KV cache to choose an element for the top-$k$ set; we report values as a fraction of top-$k$. This informs of the \textit{virtual} size of the KV space for next token prediction
  \item \textbf{New lookup count}: given the previous top-$k$ selection, how many new KV entries enter the next step's top-$k$; we report values as a fraction of top-$k$. The fraction of entries from the previous step that are not reused complements this to 1.0. This informs us of how good the previous token top-$k$ choice is a predictor of the next token top-$k$ choic
  \item \textbf{Inter-layer similarity}: given the top-$k$ set from the previous layer, how many KV indices overlap with the top-$k$ of the current layer? This metric indicates how predictable the current layer's selection is given the previous layer's indexer decisions
\end{enumerate}

All statistics were collected across all layers and sequences, then analyzed for mean and variance. Some histograms are shown in the next section.

\subsection{Profiling}
\label{sec:profiling}
To validate the impact of the DSA gather ineffeciency, we also performed inference on our evaluation set and benchmarked the perfromance of the memory subsystem of the GPU; on the NVidia architecture, we used the H200 and the NVidia Nsight Compute (NCU) performance counters.

We measured performance only in the decode stage; since NCU repeats each kernel to measure performance, we made sure the flush the L2 cache and measure the uncached performance; in any form of naive implemetation of DSA, the L2 cache is not expected to return any hits, as inference is evaluated layer by layer and token by token, and each of these fills the L2 with a completely novel subset of KV entries. 

We aggregated the results and measured the effective HBM throughput during the SDPA kernel execution. Since SDPA in decode is memory constrained, we expect high utilization, and any degradation of the utilization implies inefficiency. The results, as shown in Table~\ref{tab:profile_results}, clearly show an underutilized GPU in the sparse attention implementation.

\begin{table}[t]
  \centering  
  \caption{Resource utilization for dense vs. sparse attention as measured using NVidia NCU on H200}
  \label{tab:profile_results}
  \begin{tabular}{lcc}
    \toprule
    \textbf{Resource} & \textbf{Dense} & \textbf{Sparse} \\
    \midrule
    SM Utilization & 23.5\% & 10.9\%  \\
    HBM BW Utilization & 36.0\% & 7.3\%  \\
    \bottomrule
  \end{tabular}
\vspace{-0.75\baselineskip}
\end{table}

\section{Analysis Results}
\label{sec:analysis-results}
We analyzed the access patterns of DSA into the KV storage to gather insights on caching, management, and scaling of decode-optimized inference systems generating tokens with DSA-dominated backbones.

As stated in Section~\ref{sec:lightning-indexer}, the experimental setup was based on a knowledge-distilled open-source model and examined sequences of 500--1500 tokens each, generating 200 tokens per sequence, and using top-$k$ values of 64, 128, and 256. Fifty evaluation sequences were synthesized by an LLM to (i) span multiple domains, (ii) require long responses, and (iii) be easy to verify for answer correctness; we used a judge LLM with a weighted score combining answer correctness (given the original prompt) and grammatical correctness.

\begin{table}[t]
  \centering
  
  \caption{Analysis of access patterns to the KV storage}
  \label{tab:access_analysis}
  \begin{tabular}{lcccc}
    \toprule
    \textbf{Metric} & Unit & \textbf{Mean} & \textbf{P95} & $\boldsymbol{\sigma}$ \\
    \midrule
    Working Set & top-k & 5.15 & 7.2 & 1.02 \\
    Persistence & steps & 1.82 & 5 & 2.86  \\
    Lookback steps & top-k & 3.29 & 7.77 & 2.48 \\
    New lookups & top-k & 0.55 & 0.9 & 0.22 \\
    Inter-layer & top-k & 0.36 & 0.63 & 0.16 \\
    \bottomrule
  \end{tabular}
\vspace{-0.75\baselineskip}
\end{table}

\subsection{Working Set Size}
\label{sec:working-set-size}
\begin{figure}[h]
  \centering
  \includegraphics[width=0.4\textwidth]{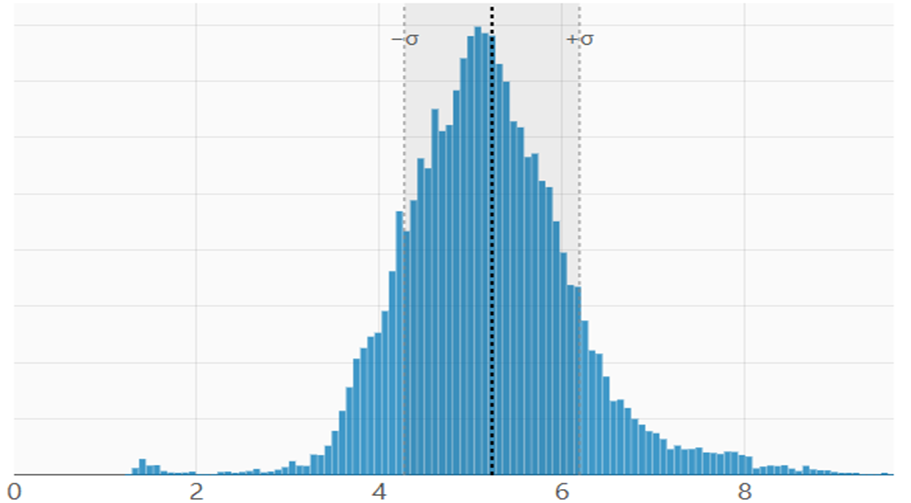}
  \caption{Distribution of the number of KV entries in a working set of a 50 consecutive tokens, as a fraction of top-k}
  \label{fig:working-set-histogram}
\end{figure}

The working set for a token chunk of size $N$ is evaluated as

\begin{equation}
  \mathcal{S}=\frac{1}{M} \sum_{m}\big(\big| \bigcup_{t=m}^{m+N} \Omega_t \big| \big)
\end{equation}

where $\Omega_k$ is the top-$k$ set at step $k$; statistics are collected across all sequences, models, and layers. The results are shown in Figure~\ref{fig:working-set-histogram} and follow an approximately normal distribution with a mean of about top-$k \times 5$ and a P95 of top-$k \times 7.2$. For example, with top-$k=128$, the working-set P95 would be $922$ KV entries per layer, per tenant. In our evaluations we used $N=50$.

\subsection{KV Entry Persistence}
\label{sec:kv-entry-persistence}
KV entry persistence is the number of time steps a KV entry remains in a specific layer's top-$k$ set for a given tenant. Our analysis shows a clear exponential falloff, with a mean of fewer than 2 time steps.

The results clearly show the spurious nature of the indexer's top-$k$ selection, and the difficulty in any useful zero-order approximations for hot KV cache region in the next decode steps.

\begin{figure}[h]
  \centering
  \includegraphics[width=0.35\textwidth]{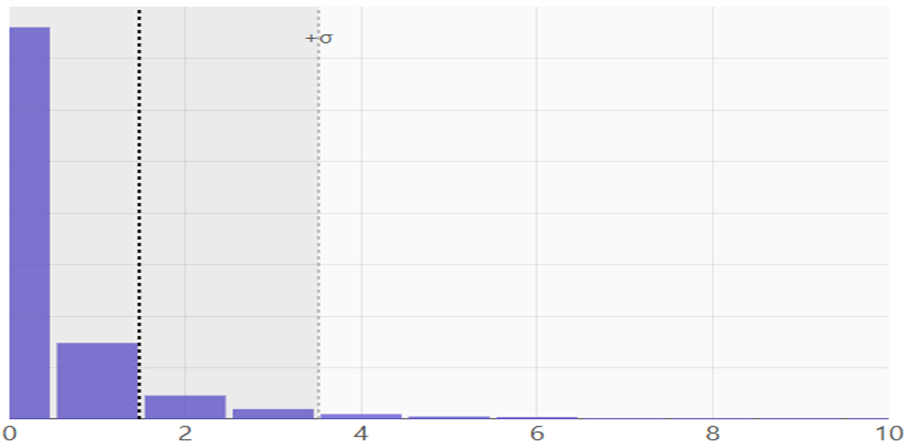}
  \caption{Distribution of persistence over time steps}
  \label{fig:persistence-histogram}
\end{figure}

\subsection{Lookback Steps}
\label{sec:lookback-steps}
Lookback measures how far into the KV cache the top-$k$ selection reaches to fetch a relevant KV entry. As we discuss below, this can inform memory tiering and hot/warm/cold labeling of KV pages.

The results suggest that lookback is mainly focused around the most recent tokens, as expected, but also with a long tail reaching out to the entire sequence in a small but significant way.

\begin{figure}[h]
  \centering
  \includegraphics[width=0.4\textwidth]{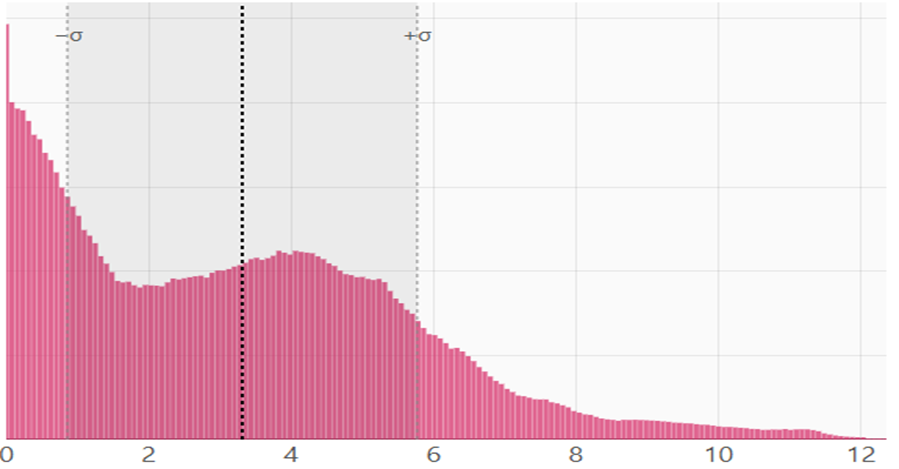}
  \caption{Distribution of lookback distance, over fraction of top-k}
  \label{fig:lookback-histogram}
\end{figure}

\subsection{New Lookup Count}
\label{sec:new-lookup-count}
New lookups are defined as the KV entries present at time step $t$ that were not present at time step $t-1$:

\begin{equation}
  L = \big| \Omega_t - \Omega_{t-1 } \big|
\end{equation}

As with the other metrics, this is averaged across sequences, token positions, and layers. The analysis shows that well over $50\%$ of the top-$k$ set at each iteration is new, making the previous time step a poor predictor of the next one.

\begin{figure}[h]
  \centering
  \includegraphics[width=0.35\textwidth]{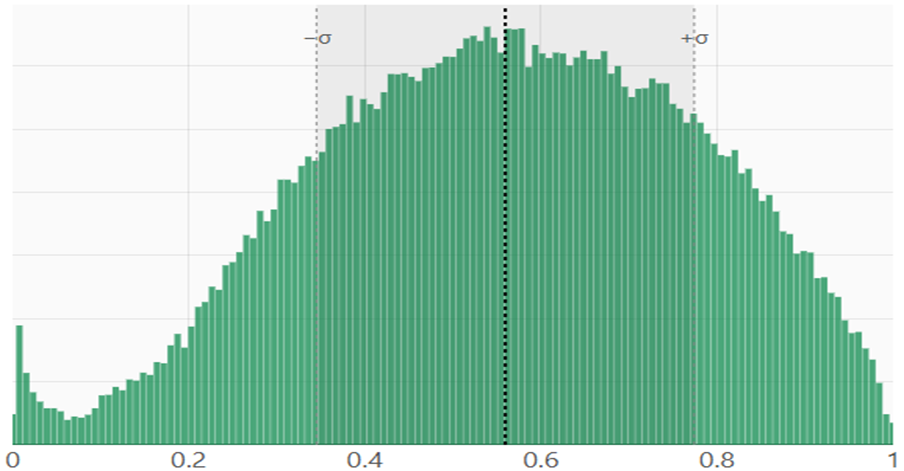}
  \caption{Distribution of new lookup count, over fraction of top-k}
  \label{fig:newLookup-histogram}
\end{figure}

\subsection{Inter-layer Index Overlap}
\label{sec:inter-layer-index-overlap}

Inter-layer index overlap is the fraction of KV \textit{indices} shared between consecutive layers. This tests the hypothesis that a predictor for the next layer's top-$k$ set can be constructed from the previous layer's selection. The results, however, negate this hypothesis: the previous layer is a very poor predictor of the next layer, with less than 50\% overlap between layers.

\subsection{Inter-layer Variations}
We analyzed several of the metrics across the different layers of the model and showed that though the values fluctuate, there are very few layers that exhibit more regular and predictable top-$k$ set selection than the average.

\begin{figure}[t]
  \centering
  \includegraphics[width=0.45\textwidth]{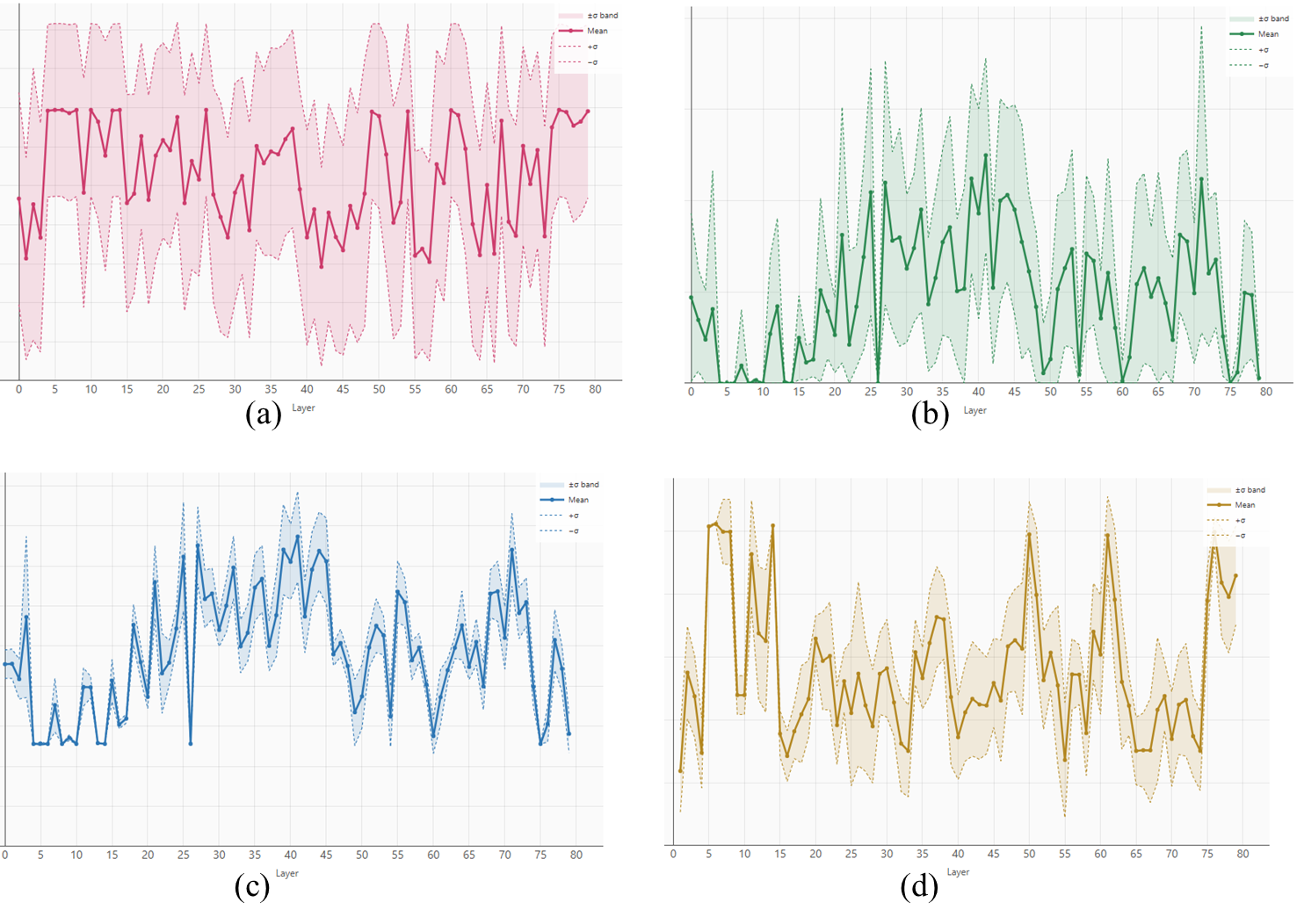}
  \caption{Mean and standard deviation of metrics across the 80 layers of LLama 3.1-70B; (a) - lookback, (b) - new lookups, (c) - working set, (d) - inter-layer overlap}
  \label{fig:cross-layer}
\end{figure}

\section{KV-granular Last Level (LL) Cache}
\label{sec:improving-l2-cache}
With the deep understanding of the access patterns that DSA imposes on the KV storage, we can propose some architectural insight on implementation of future inference devices that can be optimized for the decode stage of backbone architectures implementing DSA-like sparse attention.

As a preliminary, we note that modern GPU architectures are optimized for prefetching large tensors from HBM into L2 cache and retaining them to exploit data locality.

HBM latency is significant, often exceeding 200\,ns for each new access request \cite{asifuzzaman2021demystifying}; our assessment must take this into account. When accessing large tensor arrays, hardware prefetchers tend to \textbf{prefetch} contiguous memory pages from HBM into L2, amortizing HBM latency over the size of the fetched region. Under DSA, however, this optimization can fail because there is no guarantee that the top-$k$ set is contiguous.

Following our analysis, we conclude that the most dominant factor negatively impacting resource utilization in the DSA setting is the multiple, distinct KV page fetches as a result of the disparate \texttt{gather()} operation before computing SDPA.

Therefore we propose a simple, hardware-based architectural element that allows for allocation of a fraction of the last level (LL) cache - in NVidia architecture, that would be the L2, while in Intel architecture in would be the L3 - which will be managed as a \textit{fully associative} KV-token level management. The LRU logic fully-associative logic is usually considered expensive, however there are a relatively small number of pages (a few 1000s) in the LL cache, and the cache evaluation and eviction algorithm can take 10-20 cycles to compute, and still be an order of magnitude more efficient than a LL cache miss which would go to the main DRAM memory (HBM or LPDDR).

\begin{figure}[t]
  \centering
  \includegraphics[width=0.45\textwidth]{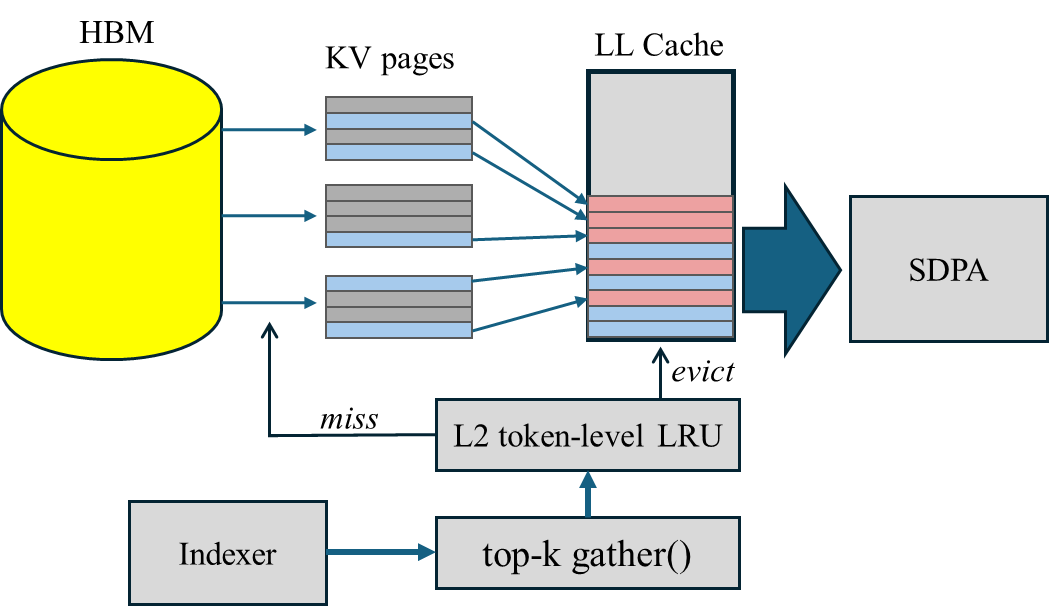}
  \caption{Proposed Architecture of KV-Granular Last Level Cache}
  \label{fig:proposed-arch}
\end{figure}

\begin{table}[t]
  \centering  
  \caption{Roofline gain for implementing LRU cache reservation for KV tokens; LLaMa 3.1-70B, 20 layers on each GPU}
  \label{tab:slowdown_analyze}
  \begin{tabular}{lccccc}
    \toprule    
    LL reserved & 0 & 5MB & 10MB & 15MB & 20MB \\
    Slowdown & 1.87 & 1.67 &  1.5 & 1.3 & 1.15 \\   
    \bottomrule
  \end{tabular}
\vspace{-0.75\baselineskip}
\end{table}

Table~\ref{tab:slowdown_analyze} shows a roofline model of how reserving a small part of LL cache and implementing an LRU KV-token mechanism could dramatically reduce system slowdown due to LL cache misses. In this example we used a 70B model, assuming 100~tokens/s/user with a batch size of 8 and 4 GPUs in parallel, so that each GPU computes ~20 layers.

The slowdown (relative to fetching the entire top-k chunk in one, efficient, HBM read) is estimated assuming each cache miss costs a delay of 200\,ns; this accumulates across layers and the batch because waiting for pages from the indexer is on the compute critical path. Our architectural modelling takes into account a paged attention approach, where batched fetches of KV tokens which reside in the same KV page only cost \textbf{one} cache miss - the most optimized possible solution (roofline).

\section{Additional Architectural Directions for Future Work}
\label{sec:dsa-impact-on-architecture}

\subsection{DSA and paged attention}
\label{sec:dsa-and-paged-attention}
\begin{figure}[h]
  \centering
  \includegraphics[width=0.35\textwidth]{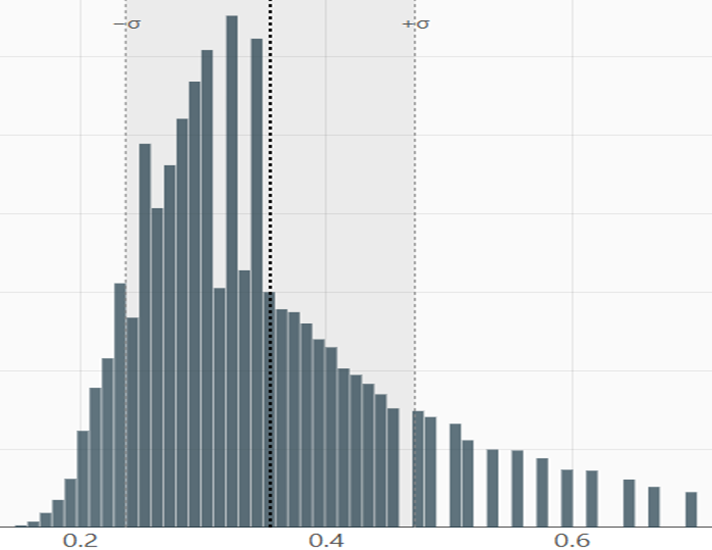}
  \caption{Distribution of average KV page utilization per decode step}
  \label{fig:pageUtil-histogram}
\end{figure}

Many modern LLM serving frameworks utilize PagedAttention \cite{kwon2023pagedattention} or variants of it. However, like other optimizations we covered, paged attention assumes that the likelihood of using any $N$ consecutive KV tokens is high; this assumption breaks under DSA. As shown in Figure~\ref{fig:pageUtil-histogram}, the average fetched KV page is only 35\% utilized; therefore, paged-attention implementations need new assumptions:
\begin{itemize}
  \item KV pages can be generated sequentially, but when read - internal fragmentation must be handled gracefully
  \item Each KV page stored in the L2 cache is partial, and a mechanism that allows this must be implemented
\end{itemize}

\subsection{Batch fetching}
\label{sec:batch-fetching}
A clear requirement that naturally proceeds our analysis, is the need for a batch fetching mechanism which is hardware accelerated; this mechanism should accept a list of access requests (reads) from HBM and be able to sort them into the correct bank access and make as many of them as concurrent as possible. This will allow a new vector for ammortization of HBM latency and fast filling of the L2 cache with the required KV values as soon as they are available from the indexer.

\subsection{Prediction of top-k}
\label{sec:prediction-of-top-k}
An inviting optimization is to predict the top-$k$ values ahead of when they are needed, to hide the latency of bringing these pages from HBM behind computation. We attempted a learned approach, training a predictor that uses hidden states from previous tokens. However, we achieved results only slightly better than keeping the previous step's top-$k$ in memory, essentially failing at this approach. We interpret this as an indication that DSA's fine-grained top-$k$ selection is highly volatile and strongly influenced by the current-step query. Other approaches to predicting top-$k$ may still be promising, but we have not explored them.

\subsection{Memory tiering of KV storage}
\label{sec:memory-tiering-of-kv-storage}
It is common practice to consider memory tiering when categorizing KV pages as hot, warm, or cold. An optimized serving framework must consider bandwidth and latency, the access patterns in each tier, and how to allocate dynamic KV storage and move it between tiers. A survey of common methods to optimize KV storage may be found in \cite{li2024survey}. Our measurements support informed decisions about which parts of the KV cache should be considered hot (very likely to be indexed in the coming chunk), warm (still likely but at much lower probability), or cold.

Our proposal in Section~\ref{sec:improving-l2-cache} is one example of this, but other directions could be explored of how to better design the architecture and management of the memory tiering to adapt to this unique workload.

\section{Conclusions}
\label{sec:conclusions}
Dynamic sparse attention (DSA) promises to lower the per-token cost of attention by limiting computation to a top-$k$ subset of KV entries, but it simultaneously shifts the decoding bottleneck toward the memory system by inducing sparse, rapidly changing KV access patterns. In this paper, we implemented a DSA-style indexer on multiple open-source backbones and instrumented per-layer KV selections during autoregressive decoding. The resulting measurements quantify (i) how quickly the effective KV working set expands over short token chunks, (ii) how little persistence individual KV indices exhibit, (iii) how far selections typically reach back into the cache, (iv) how large the step-to-step turnover is in the top-$k$ set, and (v) how weak inter-layer overlap can be. Together, these statistics explain why naive caching and locality assumptions are often insufficient for DSA and why decode throughput can degrade due to frequent L2 misses and long-latency HBM accesses.

Based on this characterization, we derived system-level implications and optimization opportunities for decode-centric inference platforms. In particular, our results motivate possible new LL cache management strategies that keep a certain subset of the KV cache within the SRAM caches, even between decode steps. We built a roofline architectural model of this proposal and combined it with the traces to show potential for tangible speedup. More broadly, the analysis suggests that future accelerators and serving stacks should co-design sparse-attention mechanisms with memory hierarchy behavior in mind, exposing primitives for KV-page tracking, selective retention, and overlap of data movement with compute.

There are several promising directions for future work. First, incorporating online or adaptive predictors for KV selection could enable more effective prefetching and reduce miss rates under distribution shift. Second, exploring possible algorithmic approach to speculatively predict the top-k set before it is needed, including usage of state-space models and other novel architectures. Third, extending the analysis to multi-tenant batching and paged KV allocation policies would clarify how fragmentation interacts with DSA under realistic serving workloads. Finally, validating these findings on hardware (or cycle-accurate simulation) and exploring hardware support for sparse, irregular gathers from KV memory tiers would help translate the measured access statistics into end-to-end throughput improvements.

\bibliography{example_paper}
\bibliographystyle{icml2026}




\end{document}